# Optogenetic manipulation of neural activity in *C. elegans*: from synapse to circuits and behavior

Steven J. Husson[1,2], Alexander Gottschalk[3], Andrew M. Leifer[4*]

[1] Functional Genomics and Proteomics, Department of Biology, KU Leuven, Naamsestraat 59, B-3000 Leuven, Belgium

[2] SPHERE - Systemic Physiological & Ecotoxicological Research, Department of Biology, University of Antwerp, Groenenborgerlaan 171/U7, B-2020 Antwerp, Belgium

[3] Buchmann Institute for Molecular Life Sciences, Johann Wolfgang Goethe-University Frankfurt, Max von Laue Str. 15, D-60438 Frankfurt, Germany

[4] Lewis-Sigler Institute for Integrative Genomics, Princeton University, 170 Carl C. Icahn Lab, Princeton, New Jersey, 08544, USA.

* To whom correspondence should be addressed: leifer@princeton.edu Phone: (609) 258-2973 Fax: (609) 258-8020

**ABSTRACT**

The emerging field of optogenetics allows for optical activation or inhibition of neurons and other tissue in the nervous system. In 2005 optogenetic proteins were expressed in the nematode *C. elegans* for the first time. Since then, *C. elegans* has served as a powerful platform upon which to conduct optogenetic investigations of synaptic function, circuit dynamics and the neuronal basis of behavior. The *C. elegans* nervous system, consisting of 302 neurons, whose connectivity and morphology has been mapped completely, drives a rich repertoire of behaviors that are quantifiable by video microscopy. This model organism's compact nervous system, quantifiable behavior, genetic tractability and optical accessibility make it especially amenable to optogenetic interrogation. Channelrhodopsin-2 (ChR2), halorhodopsin (NpHR/Halo) and other common optogenetic proteins have all been expressed in *C. elegans*. Moreover, recent advances leveraging molecular genetics and patterned light illumination have now made it possible to target photoactivation and inhibition to single cells and to do so in worms as they behave freely. Here we describe techniques and methods for optogenetic manipulation in *C. elegans*. We review recent work using optogenetics and *C. elegans* for neuroscience investigations at the level of synapses, circuits and behavior.

**Introduction**

The combination of *C. elegans* and optogenetics is a powerful platform for neuroscience investigations. The *C. elegans* model organism provides a compact nervous system of 302 neurons, whose connectome has been mapped entirely (White et al., 1986) and is capable of generating rich quantifiable behaviors including chemotaxis (Ward, 1973), thermotaxis (Hedgecock and Russell, 1975), motor sequences (Croll, 1975), habituation and simple forms of associative learning (Zhang, 2005). Optogenetics allows for the non-invasive optical manipulation of activity in neurons or other tissue. The nematode's transparent body, genetic tractability, and the consistency of neural morphology from one worm to the next, make it especially amenable to optogenetic manipulation. As such, *C. elegans* was one of the first multicellular organisms used for optogenetic experiments *in vivo* and it continues to be both a test bed for the latest optogenetic techniques as well as a popular platform for probing the nervous system at length scales spanning from synapse to circuit.

**An optogenetics toolbox for *C. elegans***

*Photostimulation of excitable cells*

Neuronal activity can be manipulated at the millisecond timescale by expressing the light-activated depolarizing cation channel Channelrhodopsin-2 (ChR2) and subsequently illuminating it with blue light (see Fig. 1). ChR2 is endogenous to the green alga *Chlamydomonas reinhardtii* and its photoactivity was first observed in oocytes from *X. laevis* (Nagel et al., 2003). Two years later, ChR2 was used to induce spiking in cultured mammalian neurons (Boyden et al., 2005). Later that year, *C. elegans* became the first multicellular organism to have its behavior manipulated by channelrhodopsin (Nagel et al., 2005). In that experiment, worms expressing ChR2 in either body wall muscles or mechanosensory neurons were induced to contract their muscles or reverse, respectively, upon illumination. In that work, ChR2's activity in *C. elegans* was also characterized with whole-cell voltage clamp recordings. An enhanced mutant of ChR2 was engineered by altering the histidine at position 134 to an arginine residue (H134R) to maximize depolarization effects. The H134R mutation results in higher peak and steady state currents with opening and closing kinetics similar to wild-type, as measured in cultured cells (Nagel et al., 2005). The ChR2 mutant H134R is the version commonly used today, and unless otherwise specified, it is this mutant that is referred to throughout the text. ChR2 has since been used to activate motorneurons, interneurons and muscles in the worm. Interestingly, the very first optogenetic experiments, performed in the lab of Gero Miesenböck, did not use ChR2 at all but instead photoactivated cultured neurons using alternative approaches such as reconstituted components of the *Drosophila* photoreceptor cascade (Zemelmann et al., 2002) or heterologous channels that were phototriggered by uncaging orthogonal ligands (Zemelmann 2003). The latter system was also used in the first-ever optogenetic experiment in a living animal, i.e. in *Drosophila* (Lima, Cell 2005). However, as these approaches are more complicated, ChR2 became more widely used.

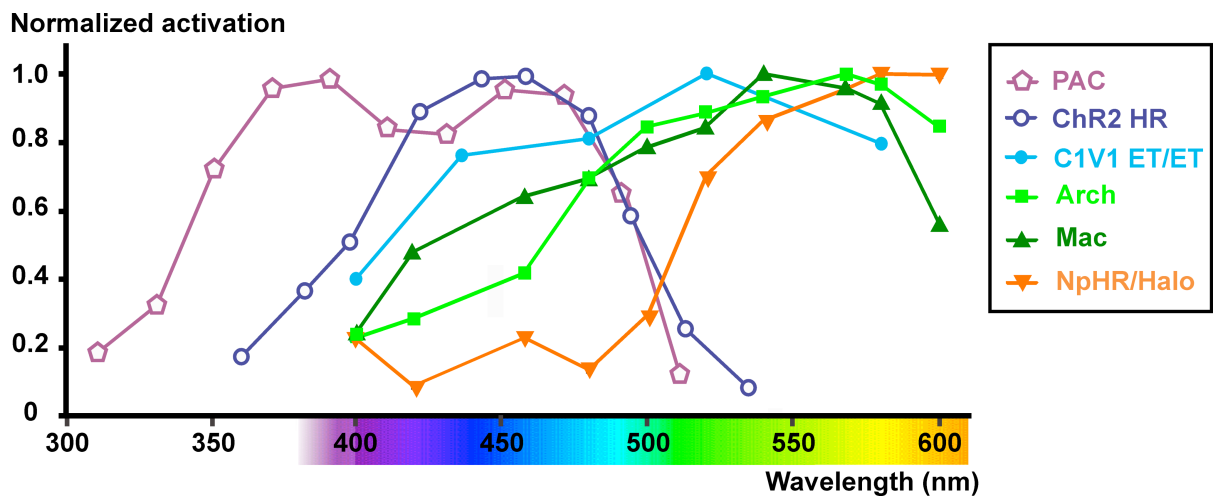

*Figure 1:* Normalized excitation spectra of optogenetic proteins used in *C. elegans*, curated from the literature. The spectrum for ChR2 is from (Zhang et al., 2007); Mac, Arch and NpHR/Halo are from (Husson et al., 2012b); C1V1 from (Erbguth et al., 2012); and PAC is from (Yoshikawa et al., 2005).

*ChR2 Variants*
ChR2 has an excitation spectrum similar to GFP with an excitation peak at 450-460 nm (see Fig. 1). It activates with sub-millisecond timescales and deactivates with timescales of order 10 ms. Other ChR2 variants provide different spectral or kinetic properties. VChR1 and C1V1 have excitation peaks in the green near 540 nm (Yizhar et al., 2011) and have been expressed in other organisms. A chimera of Chlamydomonas ChR1 and Volvox ChR1 with two point mutations (E122T; E162T), A chimera of Chlamydomonas ChR1 and Volvox ChR1 with two point mutations (E122T; E162T), called C1V1-ET/ET, is a similarly red-shifted variant (see Fig. 1) that has been expressed in *C. elegans* (Erbguth et al, 2012).

Although ChR2 has order millisecond timescale kinetics, slower variants are now also available that have deactivation time constants ranging from seconds to minutes. In particular, ChR2(C128X) mutants provide deactivation time constants of 2 s, 56 s or 106 s for mutations T, A, or S, respectively (Berndt et al., 2009) and have all been expressed in worms, allowing neural manipulations on the timescale of development (Schultheis et al., 2011a). These ChR2(C128X) variants are referred to as step-function opsins (SFOs) because, in spiking mammalian neurons, they can be used to switch a cell to an "on-state" by long-term weak depolarization. This brings the resting potential closer to action potential threshold and allows the neuron to respond to intrinsic activity more readily. Yellow light illumination inactivate these SFOs and thus can be used in combination with blue light to step neural activity on or off at arbitrary time points.

*Photoactivated adneylate cyclase (PAC)*

The optogenetic proteins discussed so far all modulate membrane potential by adjusting the flow of ions across the cell membrane. Synaptic vesicle release, however, can also be optically manipulated through intracellular second messengers acting independent of the membrane potential. Photoactivated adenylyl cyclase (PAC), first isolated from *Euglena gracilis* (Iseki et al., 2002), has been shown to manipulate the intracellular concentration of the second messenger cAMP (Schroder-Lang et al., 2007). In *C. elegans*, the PAC subunit PACα has been expressed in motor neurons. Upon photoactivation the worm's body bending increases in frequency, and the frequency of miniature postsynaptic currents in muscle also increases (Weissenberger et al., 2011). Importantly, PAC activation does not override the neurons' intrinsic activity patterns, as ChR2 activation does, but rather enhances them. One should keep in mind, though, that increasing cAMP is likely to have pleiotropic effects, as it is a second messenger involved in numerous processes in the cell. Nonetheless, tools like PAC provide an avenue with which to optically manipulate intracellular, cell biological processes, and properties of the cell that are distinct from changes in membrane potential.

*Photoinhibition of excitable cells*

The yellow-light gated $Cl^-$ pump Halorhodopsin (NpHR/Halo) from *Natronomonas pharaonis* was the first optogenetic protein shown to inhibit neural activity (Zhang et al., 2007; Han and Boyden, 2007). NpHR/Halo is the most prevalent tool for inhibiting neural activity in the worm (Zhang et al., 2007; Liu et al., 2009; Kuhara et al., 2011; Leifer et al., 2011; Busch et al., 2012, Piggott et al, 2011). When NpHR/Halo is expressed in muscles, yellow green light causes the worm's body to extend. Two other more recently discovered photoinhibitory membrane proteins (Chow et al., 2010) have also been expressed in worms: archaerhodopsin-3, known as Arch, from *Halorubrum sodomense* (Husson et al., 2012b; Okazaki et al., 2012); and Mac, from the fungus *Leptosphaeria maculans* (Stirman et al., 2011; Husson et al., 2012b). These outward-directed proton pumps have different spectral properties (see Fig. 1), and higher inward currents compared to NpHR/Halo (Chow et al., 2010; Husson et al., 2012b).

*Choosing an Optogenetic Protein*

In selecting an optogenetic protein, one must consider the protein's excitation spectra, kinetics, expression levels and the strength of the perturbation it induces, whether ionic or otherwise. Spectra considerations are especially important when expressing combinations of optogenetic proteins. A summary of excitation spectra of optogenetic proteins is shown in Fig. 1. For example, ChR2 and NpHR/Halo can be expressed together in a single neuron which can then be independently activated or inhibited with blue or green light because the two protein's excitation spectra are sufficiently narrow and non-overlapping (Zhang et al, 2007). Spectral considerations are also important when optogenetic proteins are used with optical probes in the same cell. For example, despite potential workarounds (Guo et al., 2009), it remains challenging to use ChR2 with the calcium indicator GCaMP3 because both have similar excitation spectra and thus the light required to excite GCaMP3 also activates ChR2.

The kinetics of an optogenetic protein should also be suited to the experiment. For example, ChR2 variants with slow off-kinetics are particularly useful for long-term manipulations of developmental pathways. In one such developmental experiment, ChR2(C128S) was used to rescue constitutive dauer (Daf-c) *daf-11* worms by depolarizing the ASJ neurons repeatedly over long time scales (Schultheis et al., 2011a). Additionally, not all optogenetic proteins have the same plasma membrane expression levels. Mac and Arch were recently shown to have higher photocurrents in *C. elegans* than NpHR/Halo, likely due to more efficient trafficking to the plasma membrane (Husson et al., 2012b).

Just as the number of genetically encoded fluorescent proteins rapidly increased following the introduction of GFP almost two decades ago, so too do we suspect to see a similarly rapid increase in the number and diversity of optogenetic proteins. Future researchers will likely have a wide array of optogenetic proteins with different spectral, temporal properties, expression levels and ionic specificities and conductances to choose from.

**Nuts and Bolts of Optogenetics in *C. elegans***

*Expressing Optogenetic Proteins in* C. elegans
Optogenetic proteins are expressed in *C. elegans* under the control of a promoter sequence. Thousands of promoters and their expression patterns have been annotated and are publicly searchable on WormBase (http://www.wormbase.org). Transgenic animals are generated by injecting a plasmid into the worm's gonad, or by other methods, e.g. microparticle bombardment. Transgenic lines can be frozen indefinitely. Strains, including those expressing optogenetic proteins, are available for a nominal fee from the *C. elegans* Genetics Center (http://www.cbs.umn.edu/cgc). For a review of the many genetics tools available for *C. elegans*, see the recent review (Boulin and Hobert 2012).

Certain promoters are convenient for eliciting an obvious and robust behavioral response under whole-worm illumination. These are useful when assessing the efficacy of a new optogenetic protein or when first experimenting with optogenetics. For photostimulation, the *mec-4* promoter is convenient. *mec-4* drives expression in six "gentle-touch" mechanosensory neurons. Upon photoactivation, these sensory neurons evoke an escape response whereby the worm reverses and reorients in an "omega-turn" before reinitiating forward locomotion (Nagel et al., 2005). For photoinhibition via NpHR/Halo, or for evoking spastic paralysis via ChR2, it is convenient to use *myo-3* or *unc-17* promoters which drive expression in muscles or in cholinergic motorneurons, respectively. Photoinhibition or activation of either set of cells causes the worm to paralyze either flaccidly or spastically (Zhang et al., 2007).

Promoters are available that drive expression in almost any conceivable cell type in *C. elegans*. However, there are very few promoters that drive expression in only a single cell. To optogenetically manipulate a single cell, more sophisticated genetic or optical techniques are requires, as discussed below.

*Worm Care*

Transgenic worms expressing optogenetic proteins are grown on agar plates using standard techniques (Brenner 1974) with minor modifications. The optogenetic proteins require the cofactor *all-trans* retinal (ATR). Since nematodes do not generate ATR, exogenous ATR is added to the bacteria lawn that serves as the worm's food (Nagel et al., 2005). Transgenic worms grown without ATR serve as convenient negative controls for optogenetic experiments. For general *C. elegans* methods, see especially the online Wormbook (http://www.wormbook.org).

*Illumination*

Illumination from a standard fluorescent microscope is sufficient to induce a behavioral response in transgenic worms. A mercury lamp filtered by a GFP excitation filter producing ~1 mW/mm$^2$ of blue light (450–490 nm) will induce an escape response in worms expressing ChR2 under the *mec-4* promoter (Nagel et al., 2005). Other optogenetic proteins expressed in *C. elegans* are commonly activated with light intensities in the 0.5 to 5 mW/mm$^2$ range. Importantly, wild-type *C. elegans* also has an intrinsic photophobic response. Blue, violet, and particularly UV light can be toxic to the worm at high intensity (Edwards et al., 2008) and can induce even wild-type worms to reverse. There are a number of strategies to avoid this response. Short light pulses seem to avoid the photophobic response. Additionally, *lite-1* mutants lack the photophobic response entirely (Edwards et al., 2008) and are sometimes used instead of a wild-type background. This should be avoided when possible, however, because the *lite-1* animals appear less healthy and have a different swimming gait than wild-type. For experiments requiring long-term photoactivation, "slow" ChR2 variants discussed above offer another alternative.

Early optogenetic experiments in *C. elegans* illuminated the entire worm with the result that every cell expressing an optogenetic protein was activated. Studying the contribution of individual neurons requires either finding single-cell promoters or generating patterned illumination targeted to single-cells.

*Expressing proteins in single cells lacking single-cell promoters*

Despite the large library of known promoters, finding promoters for single-cell expression remains a challenge. In some cases, existing promoters can be split into smaller promoter fragments that then can drive expression in smaller subsets of cells. Another approach is to use a combinatorial genetic approach to express ChR2 only in cells at the intersection of two promoters. This strategy combines overlapping sets of promoters with specific transgene structures and the use of a recombinase like FLP or Cre (Davis et al., 2008; Macosko et al., 2009; Schmitt et al., 2012) (see Fig. 2). In this approach, a first promoter is cloned in front of an optogenetic protein of interest. This construct, however, contains a transcription termination sequence that is flanked by recombinase recognition sequences (e.g. *loxP* or *FRT*) sites. A second promoter drives expression of the recombinase (Cre or FLP). The enzymatically mediated recombination of these recognition sites removes the transcriptional stop sequence and enables expression in only cells where both promoters are active (Schmitt et al., 2012). Sub-populations of neurons, or even single cells, can be further addressed by restricting the illumination pattern to the cell of interest, as explained in the next section.

*Targeted Illumination*

In contrast to whole-field illumination, targeted illumination complements genetic specificity by providing spatial specificity. If two neurons express an optogenetic protein, each neuron can be photoactivated independently by shining light on only one or the other, provided that the two are sufficiently far apart. In its simplest form, targeted illumination requires only a standard fluorescent microscope with a high-magnification objective and an aperture in a conjugate plane to the specimen plane that can be used to restrict excitation light to a small spot. An immobilized worm can then be positioned such that the spot illuminates only the cells of interest.

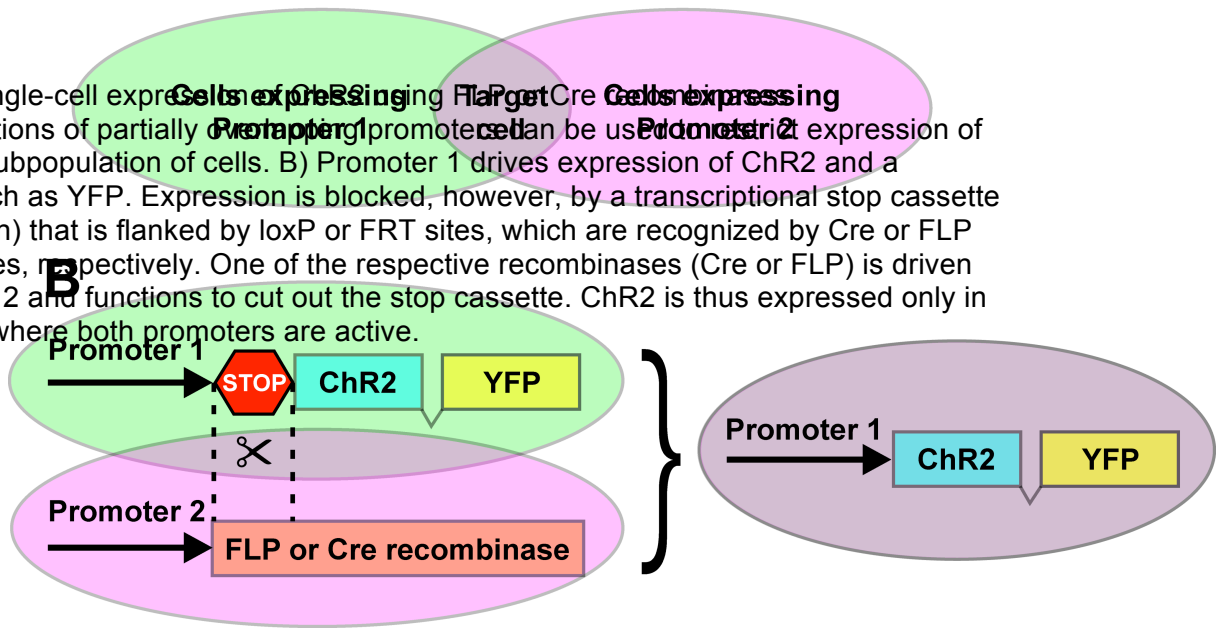

Figure 2: Single-cell expression of ChR2. A) Combinations of partially overlapping promoters can be used to restrict expression of ChR2 to a subpopulation of cells. B) Promoter 1 drives expression of ChR2 and a reporter, such as YFP. Expression is blocked, however, by a transcriptional stop cassette (red hexagon) that is flanked by loxP or FRT sites, which are recognized by Cre or FLP recombinases, respectively. One of the respective recombinases (Cre or FLP) is driven by promoter 2 and functions to cut out the stop cassette. ChR2 is thus expressed only in target cells where both promoters are active.

Spot illumination is sufficient for single cells or contiguous cells, but to illuminate multiple cells in distinct locations or to rapidly switch between targeted cells requires patterns of illumination and the ability to spatio-temporally modulate such patterns. Guo et al. were the first to use spatio-temporally patterned illumination with immobilized *C. elegans*. They used a digital micromirror device (DMD) to activate ChR2 in neurons including the polymodal sensory neuron ASH while optically monitoring calcium levels in other neurons (Guo et al., 2009). DMDs are commonly found in high-end digital projectors and consist of hundreds of thousands of microscopic independently addressable mirrors that can be adjusted to reflect arbitrary illumination patterns onto the worm. In principle, any patterned illumination system can be used to illuminate worms including galvanometer mirrors, acousto-optic deflectors (AOD), spatial-light modulators or liquid crystal display (LCD) projectors.

*Targeted illumination of freely moving worms*

It is often desirable to observe behavior while manipulating neural activity. Illuminating single cells in a moving worm requires updating the illumination pattern in response to the worm's motion in real-time. Two closed-loop systems to manipulate neural activity in freely moving worms were published simultaneously in 2011: one used an LCD projector (Stirman et al., 2011) and the other, called CoLBeRT, used a DMD (Leifer et al., 2011). The systems were used to stimulate and inhibit collections of muscles, motor neurons and even individual mechanosensory neurons in unrestrained worms while simultaneously observing behavior. Both systems take advantage of the worm's stereotyped morphology to automatically identify targeted cells based on real-time images of the worm's body as it moves. Custom computer vision software infers the location of targeted neurons by analyzing the outline of the worm's body. A detailed comparison of the two systems is provided in (Stirman et al., 2012). Briefly, the LCD projector system is less expensive, easier to setup and provides independent control of multi-color illumination, while the CoLBeRT system is more accurate. The DMD approach has been further adopted in another recent work that targets individual neurons close together in the nerve ring in moving worms (Kocabas et al., 2012).

The primary factor affecting the accuracy of these closed-loop systems is the latency between imaging the worm and updating the illumination pattern. Note that latency is not the same as frame rate. If the latency is too high, the illumination pattern is unable to keep pace with the target cell's motion and can miss the target or errantly activate an incorrect cell. The accuracy of these systems can be tested in a physiologically relevant way by illuminating cells in a worm expressing a photoconvertible protein such as Kaede. Kaede's fluorescence spectrum irreversibly changes upon activation by violet light (Ando et al., 2002). As a result, Kaede provides a record of where the worm was illuminated (Leifer et al., 2011).

**Analysis of synaptic transmission**

*C. elegans* is an important system to study neurotransmission at chemical synapses. In fact, many of the most crucial players in synaptic transmission, synaptic vesicle (SV) docking, priming, fusion and recycling were discovered first in *C. elegans* (e.g. UNC-13, UNC-18), and were later confirmed to function in a highly conserved manner in mammals (Richmond, 2005; Schuske et al., 2004; Barclay et a., 2012; Bargmann and Kaplan, 1998).

Early on, analysis of synaptic transmission genes relied mostly on pharmacological assays to infer whether a defect occurred either pre- or postsynaptically (Lewis et al., 1987; Miller et al., 1996). These assays are slow and require analyzing populations of worms. The development of electrophysiological techniques enabled the recording of neuronal activity directly, either pre- or postsynaptically, to study synaptic release of transmitter, and the receptors detecting them (Richmond, 2009; Richmond 2006; Francis et al., 2003; Francis and Maricq, 2006; Goodman et al., 2012). Electrophysiology was first used in neurons (Goodman et al., 1998), and then at the neuromuscular junction, on muscle (NMJ) (Richmond and Jorgensen, 1999). Electrophysiology experiments in *C. elegans*, however, have a number of limitations. The experiments themselves are technically challenging due, in part, to the difficult dissection required. Electrophysiology methods do not permit specific stimulation of just one receptor type, making analysis of individual receptors challenging. In electrophysiology studies of presynaptic defects, the techniques available for evoking synaptic vesicle release are also limited and some even cause neuronal damage and cannot be used repeatedly, thus hampering studies of synaptic plasticity. Finally, electrophysiology offers no method to stimulate synaptic vesicle release in intact behaving *C. elegans*.

Here, optogenetics provides clear advantages and has opened entirely new experimental possibilities. Specifically, optogenetics allows stimulation in intact moving animals, is repeatable, technically less challenging, provides specificity of stimulation for neuron type, and allows the induced release of endogenous transmitter at synapses only and in natural amounts.

The *C. elegans* NMJ comprises cholinergic and GABAergic motor neurons, and postsynaptic ionotropic and metabotropic receptors for their respective transmitters. Their function and properties are active areas of investigation. Two papers, one in 2008 (Liewald et al., 2008) and one in 2009 (Liu et al., 2009), analyzed synaptic transmission at the NMJ using ChR2-mediated photostimulation. ChR2 was expressed in cholinergic or GABAergic neurons, using specific promoters. These neurons could be specifically, reliably and strongly photoactivated. In intact animals, this evokes either contraction or relaxation of the body, which can be measured and used as readout for pre- or postsynaptic functionality. Importantly, these neuron types could also be photostimulated in dissected animals to allow the release of endogenous transmitters locally, at synaptic sites.

Liewald et al. analyzed a set of pre- and postsynaptic mutants, both by optogenetic and electrophysiological methods. When cholinergic neurons were photostimulated, presynaptic mutants paradoxically evoked stronger muscle contraction than wild type animals. This could be attributed to compensatory alteration of muscular excitability in mutants that release reduced amounts of acetylcholine. For GABAergic neurons, the effects were as intuitively expected, namely presynaptic defects led to weaker photo-evoked muscle relaxation. Behavioral experiments for GABA photostimulation further yielded evidence for a $GABA_B$ receptor acting in cholinergic neurons (Schultheis et al., 2011b). In electrophysiological experiments, clear correlation between functionality of presynaptic release machinery and amount of observed postsynaptic current was evident, both for GABAergic and cholinergic neurons. The same techniques could also be used to study postsynaptic nAChR function (Almedom et al., 2009; Liu et al., 2009). Optogenetics provided a number of additional insights into the NMJ, for example that NMJ neurons evoke transmitter release in a sustained (tonic) and graded fashion and the postsynaptic evoked current has an approximately log-linear correlation with the intensity of light stimulus (Liu et al., 2009; Schultheis et al., 2011b). It was also found that upon repeated stimulation, GABAergic neurons showed some facilitation, while cholinergic neurons showed depression (Liu et al., 2009; Liewald et al., 2008), and that NpHR/Halo activation in motor neurons suppressed transmitter release as evidenced by a disappearance of miniature postsynaptic currents. Two recent studies also probe the role of gap junctions in coordinating muscle activity and explore how neuronal transmission drives action potential generation and contractions in body wall muscle (Gao and Zhen, 2011; Liu et al., 2011).

Moreover the light activated adenylyl cyclase, PAC, has been used to manipulate cAMP production at the NMJ. Photostimulation of PAC in cholinergic neurons (Weissenberger et al., 2011) led to increased locomotion activity that was coordinated, in contrast to the spastic paralysis seen when photostimulating these neurons via ChR2. PAC photostimulation increases the rate of synaptic vesicle fusion events, possibly by promoting synaptic vesicle priming, and also causes slightly elevated amplitude of mini events. While the reason for the latter is unclear, the study demonstrates that directly manipulating membrane potential can have different effects at the synapse, compared to manipulating cAMP levels.

In addition to at the NMJ, synaptic transmission in or between neurons is also an active area of research. Electrophysiology has been the classical tool to probe spontaneous activity or synaptic transmission in or between neurons by directly recording from neurons that are carefully dissected out of the cuticle. This approach was used in a number of applications where neuronal properties were analyzed or receptor currents were measured, e.g. in thermosensory, nociceptive or mechanosensory neurons in response to natural stimuli (O'Hagan et al., 2005; Geffeney et al., 2011; Ramot et al., 2008; Kang et al., 2010; Mellem et al., 2002; Kawano et al., 2011). As these studies rely on spontaneous or naturally evoked synaptic transmission via sensory neurons, they could not be performed for other, non-sensory neurons. However, recently, these types of studies have adopted optogenetic approaches, called "photo-electrophysiology", to characterize neuron-neuron transmission without the need to use a natural stimulus. This made central synapses newly accessible to electrophysiological analysis. Synaptic transmission at a select number of sensory neurons and interneurons has been analyzed this way. One study used optogenetic stimulation to probe parameters of synaptic transmission between a thermosensory neuron, AFD, and its downstream interneuron, AIY (Narayan et al., 2011). Here, ChR2 could depolarize AFD up to 40 mV, and synaptic transmission to AIY was tonic and graded. This was probed by sustained photoactivation of AFD at light intensities spanning five orders of magnitude. Interestingly, downstream currents and depolarization in AIY were rather small, (about 2 pA and 2 mV each), and depended on peptidergic signaling. The neuropeptide and receptor responsible for transmission, however, were not identified. Another photo-electrophysiology experiment analyzed transmission between the aversive polymodal sensory neurons ASH and the backward command interneuron AVA (Lindsay et al., 2011). In that experiment, ASH was photoactivated using ChR2. Again, synaptic transmission was observed to be graded, i.e. transmission increased with increasing light intensity, and it was demonstrated that transmission depends on glutamate. When conducting these "photo-electrophysiology" experiments it is important to account for variability in ChR2 expression and for the fact that ChR2's peak current can change during prolonged or repeated stimulation (Liewald et al., 2008; Liu et al., 2009).

All *C. elegans* neurons analyzed by photo-electrophysiology thus far exhibitgraded transmission. Namely, transmitter release directly correlates with the extent of depolarization evoked by light-activating ChR2. This is in contrast to mammalian systems, for example, which exhibit action potentials, or spikes. This difference must be taken into account before applying principles of synaptic transmitter release gleaned from *C. elegans* to other organisms. Moreover, compared to spiking neurons, the graded nature of transmission in *C. elegans* has practical consequences for optogenetically inducing transmitter release. In - needs only to induce depolarization from resting potential (~65 mV) up to threshold (~5-10 mv) to activate sodium channels, trigger an action potential (~100 mV) and thus achieve maximal transmitter release. In *C. elegans* neurons, which lack voltage-gated sodium channels, one needs to externally induce depolarization without assistance from action potentials. As a result, higher induced depolarization is required to achieve maximal transmitter release. More generally, ChR2 has weak sodium conductance compared to a typical voltage-gated sodium channel. Thus, to induce sufficient depolarization for the desired level of transmission or behavior, it is often helpful to employ a combination of the following strategies: use a strong promoter that drives high levels of ChR2 expression, use high copy numbers of ChR2, use ChR2 variants that provide higher current such as the ChR2(H134R; T159C) double mutant (Erbguth et al., 2012), and adjust the illumination intensity and duration to provide brief bright light pulses that still avoid the photophobic response.

Certain types of experiments would be challenging with spiking neurons but lend themselves particularly well to graded transmission. For example, stimuli can be tuned by "titrating" light intensity to adjust the extent of ChR2-mediated current and depolarization. This way, even the extent of a behavior can be modified in a graded fashion, as exemplified by the extent of evoked escape velocity of animals in which nociceptors were photostimulated (Husson et al., 2012a; see below).

In addition to the above examples, optogenetic stimulation has also been used to study signaling in other tissues, e.g. to investigate signaling between the intestine, GABAergic neurons and enteric muscles involved in the defecation motor program (Mahoney et al., 2008). Here, photostimulation of GABAergic neurons could bypass the lack of several signaling molecules that are required to activate GABAergic neurons in this context. Last, cholinergic neurons in the pharyngeal nervous system were photostimulated using ChR2, to verify the involvement of acute cholinergic signaling in initiation of the pharyngeal action potential (Franks et al., 2009). More generally, optogenetics has played a crucial role in making stimulation and recording more accessible for investigations of signaling and synaptic transmission.

**Dissection of neuronal circuits**

An underlying goal in neuroscience is to understand how collections of neurons integrate sensory inputs to generate behavioral outputs. With its mapped connectome and compact nervous system, the nematode *C. elegans* provides a valuable test-bed for studying the function of elementary neural circuits. Here we focus on the so-called "wired network" of synapses and gap junctions and the role that optogenetics has played in probing the network's functional activity. We note, however, that important neural information likely also flows through means that extend beyond the mapped connectome, including through signaling of slow-acting transmitters such as neuropeptides or biogenic amines. *C. elegans* has been used to explore the role of neural extrasynaptic signaling in behavior, for example, see (Chase et al., 2004).

Early neural circuit experiments in *C. elegans* utilized the wiring diagram in combination with laser killing experiments and mutant analysis to infer the role of specific neurons for behavior (White et al., 1986; Sulston and White, 1980; Chalfie et al., 1985). By quantifying behavioral defects, investigators successfully identified the role that specific sensory neurons, interneurons and motor neurons played in behaviors such as mechanosensation and locomotion (Chalfie et al., 1985; Zheng et al.,1999). The models that emerged are impressive in their ability to answer the question which neurons are involved in which behavior. However, they do not attempt to address the more complicated question of what are the temporal patterns of neural activity that drive such behavior. Laser killing and genetic techniques do not allow for temporally precise and reversible perturbations to neural activity, nor do they provide a readout of circuit activity other than the worm's behavior.

Electrophysiology provides temporal precision and allows neural activity to be recorded and perturbed reversibly. However, the preparation for electrophysiology in worms requires that the worm be dissected and immobilized which severely disrupts behavior. Moreover, in the compact worm it is technically challenging to record or stimulate from more than one neuron at a time. As a result, early electrophysiology experiments primarily characterized receptors or synapses (see previous section), but not circuits.

In contrast to other techniques, optogenetics allows the investigator to reversibly induce or inhibit neural activity remotely and observe behavior, even in unrestrained worms. Optogenetics allows for the targeting of neurons otherwise inaccessible with electrophysiology, and it allows for simultaneous manipulation of multiple neurons at once. In combination with calcium imaging and other techniques, optogenetics has yielded several new insights about the role neurons play in specific circuits. Here we review examples from different circuits and systems in *C. elegans* where optogenetic techniques have played an important role in revealing underlying circuit mechanisms (see Fig. 3). For a table of optogenetic experiments conducted in *C. elegans* by neuron, circuit or system, see Table 1.

*Forward and reverse locomotion circuit*
Many of the interesting behaviors of *C. elegans*, like learning, chemotaxis, thermotaxis, mating and lethargus, manifest themselves as changes to the worm's locomotion. In particular, the worm punctuates its forward locomotion with reversals, which it uses to avoid a stimulus or to execute navigational strategies. As a result, the study of forward and reverse locomotion serves as an entry into understanding the worm's motor circuit.

The critical neurons for controlling forward and backward locomotion were identified from the worm's wiring diagram and genetic and laser ablation studies (Chalfie et al., 1985; Zheng et al., 1999). Based on these studies, the interneurons AVA, AVD and AVE are associated with reverse locomotion and interneurons PVC and AVB are associated with forward locomotion. Downstream, A-type motor neurons carry out reversals while B-type motor neurons carry out forward motion. The interneuron RIM resides in the network between these forward and reverse interneurons and is important for head bending and reversals (Alkema et al., 2005; Pirri et al., 2009).

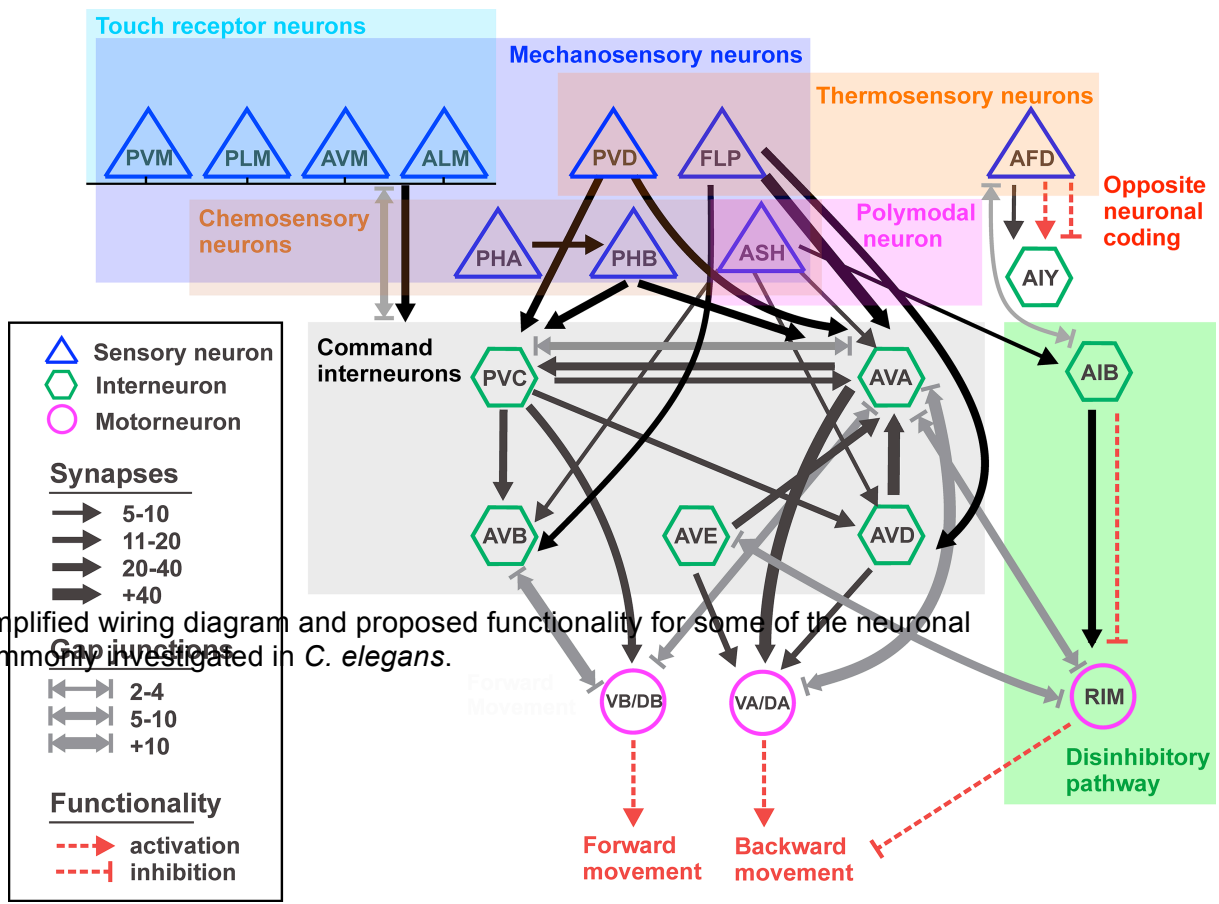

*Figure 3:* Simplified wiring diagram and proposed functionality for some of the neuronal networks commonly investigated in *C. elegans*.

While existing models provide a static picture of the motor circuit, optogenetics and calcium imaging have provided new information about functional dynamics of the circuit. Consistent with the model above, photostimulating AVA induces backward motion (Schmitt et al., 2012) and photostimulating PVC results in accelerations (Husson et al., 2012a; Stirman et al., 2011). Calcium imaging studies such as (Kawano et al., 2011) have provided additional details, showing that AVA & AVE co-activate and are anti-correlated with AVB, while the B type and A type motor neurons are responsible for forward and reverse locomotion, respectively.

| Circuit | Cells Manipulated | References |
|---|---|---|
| Forward and Reverse locomotory circuit | AVA, AVD, AVE, PVC, AVB, AIB, RIM | (Stirman et al., 2011; Piggott et al., 2011; Husson et al., 2012a; Husson et al., 2012b; Schmitt et al., 2012) |
| Motor Circuit, Wave propagation | Cholinergic motor neurons, muscles | (Nagel et al., 2005; Zhang et al., 2007; Leifer et al., 2011; Stirman et al., 2011; Husson et al., 2012b; Singaram et al., 2011; Wen et al., 2012) |
| Mechanosensation (gentle touch) | ALM, AVM, PVM, PLM | (Nagel et al., 2005; Leifer et al., 2011; Stirman et al., 2011; Timbers et al., 2013; Husson et al., 2012a; Husson et al., 2012b) |
| Mechanosensation (harsh touch) | PVD, FLP, PHA, PHB, BDU, SDQR, AQR | (Li et al., 2011; Husson et al., 2012a) |
| Polymodal nociception | ASH | (Guo et al., 2009; Ezcurra et al., 2011; Faumont et al., 2011; Lindsay et al., 2011; Husson et al., 2012b) |
| Oxygen sensation | AQR, PQR, URX | (Milward et al., 2011; Busch et al., 2012) |
| Thermosensation | AFD, AIY, AWC | (Kuhara et al., 2011; Narayan et al., 2011; Kocabas et al., 2012) |
| Copulation | A- and B-type ray neurons | (Koo et al., 2011) |
| Egg-laying | HSN | (Emtage et al., 2012; Leifer et al., 2011) |
| Proprioception | DVA | (Feng et al., 2012; Wen et al., 2012) |
| Chemosensation | AWC, AIY | (Kocabas et al., 2012) |

*Table 1:* Circuits, systems and neurons that have been interrogated using optogenetic techniques.

Optogenetic investigations especially have raised new questions about the motor circuit. One area of particular interest is the role that RIM plays. Photostimulation of RIM induces the worm to reverse (Guo et al., 2009), in agreement with previous evidence that RIM and AVA activity are correlated through gap junctions (Alkema et al., 2005) and that RIM inhibits AVB (Pirri et al., 2009). Recently however, a study combining optogenetic manipulations with other techniques suggests that RIM plays a much more complicated role (Piggott et al., 2011). In that work RIM was inhibited using NpHR/Halo and the authors observe that it - counter intuitively- induces reversals, even in animals where the reversal interneurons (AVA-AVD-AVE) had been killed. Moreover, in those animals lacking AVA-AVD-AVE, photostimulating RIM with ChR2 failed to induce reversals. The authors show that RIM plays two opposing roles in modulating reversals depending on the nature of the sensory stimulus - even when both stimuli are mediated by the same sensory neuron, ASH. The authors suggest a new parallel pathway for reversal initiation. In this model, reversals are induced when neuron AIB inhibits RIM through a pathway that could either bypass the AVA-AVD-AVE interneurons, or that could work in concert with them. The existence of this parallel pathway raises questions about redundancy and complexity in the motor circuit and suggests that a single neuron's activity can play multiple roles depending on context.

*Wave Propagation & Proprioception*

To travel forward, the worm propagates sinusoidal body bending waves from anterior to posterior. An active area of research has been to understand the mechanism by which these oscillatory bending waves are generated and propagated. Optogenetics played an important role in recent work showing that the worm propagates body bending waves by proprioceptive coupling between adjacent body regions (Wen et al., 2012). According to this model, the worm senses its own bending in anterior body segments, which reflexively induces bending in posterior body segments with a time delay. Optogenetics was used to determine which part of the motor circuit carried this proprioceptive feedback between body segments. By selectively inhibiting or exciting muscles or motor neurons in specific region of the worm, and observing how wave propagation was disrupted, the authors showed that cholinergic B type motor neurons directly sensed the body bending and transduced a proprioceptive signal.

*Neuronal circuits regulating mechanosensation: gentle and harsh touch*

The mechanosensory circuit is arguably the best characterized neural circuit in *C. elegans*. Six touch receptor neurons (TRN) detect "gentle touch" to different regions of the body, triggering forward or backward escape reflexes (Chalfie et al., 1985). ChR2-mediated photoactivation of TRNs also evokes reversals (Nagel et al., 2005), and this technique has proven useful for filling in functional details of the mechanosensory circuit. Photoactivation of anterior touch receptors induced reversals, while stimulating posterior touch receptors induced accelerations (Leifer et al., 2011; Stirman et al., 2011). Stimulating a single individual mechanosensory neuron is sufficient to induce reversals (Leifer et al., 2011) and reversals can be blocked by photoinhibition of downstream interneurons (Husson et al., 2012b; Stirman et al., 2011). Optogenetics has also allowed for a more detailed probing of habituation (Leifer et al., 2011), including the effects of aging on habituation (Timbers et al., 2013).

In contrast with gentle touch, harsh touch is sensed by the multidendritic FLP and PVD neurons (Chatzigeorgiou et al., 2010; Chatzigeorgiou and Schafer, 2011; Oren-Suissa et al., 2010; Way and Chalfie, 1989). Using laser ablations, additional cells (BDU, SDQR, AQR, ADE, PDE, PHA and PHB) contributing to harsh touch sensation could be identified (Li et al., 2011). For the first time, optogenetics allowed for the study of harsh touch circuits independent of the gentle touch receptors which would normally be co-activated by harsh mechanical stimulation. Photoactivation of single or multiple cells of the circuit were shown to induce escape responses similar to those resulting from the endogenous stimuli (Husson et al., 2012a; Li et al., 2011). Photostimulation of PVD increased calcium levels in PVC and resulted in a forward escape movement (Husson et al., 2012a). In contrast, *deg-1(d)* animals with degenerated PVC neurons robustly moved backward (Husson et al., 2012a), showing that in wild type animals the overall excitatory PVD-AVA synapses are overruled by the PVD-PVC synapses, leading to forward escape reactions. Because optogenetic stimulation of PVD bypasses endogenous mechanotransduction channels, it was also used to uncover genes required for PVD function that operate downstream of primary sensory molecules (Husson et al., 2012a).

*Dissection of the polymodal nociceptive ASH circuit*
The polymodal ASH neurons are capable of sensing diverse input signals like chemicals (Hilliard et al., 2004; Hilliard et al., 2002; Troemel et al., 1995), osmotic stress (Bargmann et al., 1990) and nose touch (Kaplan and Horvitz, 1993), resulting in a reversal that is often followed by an omega bend. Consistently, photostimulation of this nociceptor neuron recapitulates the endogenous withdrawal behaviors (Faumont et al., 2011). Additionally the probability of a reversal and omega bend depended on light intensity, and the probability of an omega bend was also dependent on the duration of the photostimulus (Ezcurra et al., 2011; Husson et al., 2012b). Optogenetics-assisted photodepolarization of ASH yielded robust calcium transients in the downstream neurons AVA and AVD, as assessed by simultaneously photoactivating the sensory neuron and imaging the command interneurons (Guo et al., 2009). Analogously, electrophysiology recordings in AVA show evoked currrents in response to nose touch (Mellem et al., 2002), and photostimulation of ASH (Lindsay et al., 2011). When Mac or Arch were used to block downstream signaling in the command interneurons, the photoevoked backward movement upon stimulation of ASH was temporarily blocked, but continued after the inhibitory light pulse (Husson et al., 2012b). The ASH neurons thus continue to signal to downstream interneurons during extended photostimulation, which is supported by elevated Ca$^{2+}$ levels in ASH neurons when stimulated for longer periods (Hilliard et al., 2005). Such physiology is appropriate for a nociceptive neuron that is expected to continue to signal the threatening stimulus as long as it is detected. Additionally, recent evidence suggests that two separate parallel downstream circuits may carry out reversals induced by ASH and that the circuit used depends on whether ASH is stimulated by nose-touch or by osmotic shock (Piggott et al., 2011).

*Chemotaxis*

The worm will navigate chemical gradients to avoid harmful chemicals and to seek out food. The neuron AWC had been implicated in chemosensation. AIY is a neuron downstream of AWC, and although previous laser ablation studies had shown it to be unnecessary for chemotaxis (Bargmann et al., 1993; Ha et al., 2010), a recent optogenetic study (Kocabas et al., 2012) showed that optogenetic manipulation of AIY alone is sufficient to recapitulate the navigational aspects of chemotaxis. The work also demonstrated the power of targeted illumination in moving animals. In this case the targeted illumination system was used to photostimulate AIY only when the worm's head swung in a particular direction, thus inducing the patterns of neural activity previously observed when the animal navigates an odorant gradient. This work provides new functional evidence of the chemosensory circuit's complexity and robustness, and is an example of "closed-loop" optogenetic stimulation based on behavior.

*Tonic signaling of oxygen sensors* AQR, PQR and URX

Tonic receptors communicate stimulus duration and intensity, hereby controlling homeostasis. Homeostatic responses mediated by the AQR, PQR and URX neurons allow *C. elegans* to escape high (21%) and low (<5%) oxygen concentrations (Cheung et al., 2005; Gray et al., 2004). Photostimulation was sufficient to mimic unfavorable oxygen levels, thereby modulating sustained signaling from AQR, PQR and URX neurons and inducing food-leaving behavior (Busch et al., 2012; Milward et al., 2011). By combining different natural stimulus dynamics and Ca$^{2+}$ imaging, molecular mechanisms underlying tonic signaling from these oxygen sensors could be identified (Busch et al., 2012).

*Thermosensation*

The worm senses temperature through the thermosensory neurons AFD and its downstream target, AIY, as well as through AWC and ASI (Mori and Ohshima, 1995; Beverly et al., 2011). A recent optogenetic study revealed that the level of AFD activation determines opposite seeking behaviors: attraction or repulsion. A systematic dose-response experiment using different temperature increments or decrements indicated that sub-maximal AFD responses induced the strongest response in the downstream neuron AIY. In contrast, the strongest [$Ca^{2+}$] increments in AFD resulted in weaker AIY responses (Kuhara et al., 2011). When pulsed NpHR/Halo excitation was used to attenuate AFD activity, higher [$Ca^{2+}$] increments could be measured in AIY. These counter-intuitive results can be interpreted by assuming that AFD transmits both excitatory and inhibitory signals to its postsynaptic target AIY. The transfer characteristics of the AFD-AIY synapse were investigated by combining optogenetics with monitoring of downstream transients using electrophysiology. Optical stimulation of AFD and patch clamp recordings in AIY revealed excitatory, tonic and graded signaling that also involves peptidergic neurotransmission (Narayan et al., 2011).

*Further integrated studies including optogenetics approaches*
Copulation is a complex behavior in *C. elegans*, involving different sub-behaviors like specific turning movements of the male worm to search for the location of the vulva, insertion of the spicule and ejaculation. The A-type ray neurons are required for all appositional postures, and their photoactivation *via* ChR2 can mimic the scanning and turning-like appositional postures (Koo et al., 2011). In contrast, B-type ray neurons are only required for the initiation of this behavior. Both neuron types also seem to evoke different ventral curl postures of the tail; however, the posture evoked by the A-type neurons dominates when co-photoactivating A- and B-type neurons. Optogenetics-induced behaviors were also assessed for different mutants to elucidate critical neurotransmitter molecules (Koo et al., 2011).

In another study, optogenetics-assisted changes of neuronal membrane potential enabled modulating sensitivity to defined anesthetics (Singaram et al., 2011). The neuronal resting membrane potential, which is co-established by leak channels selective for $K^+$, or permeable to $Na^+$, was previously suggested to control anesthetic sensitivity. Indeed, ChR2-induced depolarization of cholinergic neurons was used to reverse halotane-induced immobility, whereas NpHR/Halo-triggered hyperpolarization rendered the animals more sensitive. Further studies include the functional analysis of the homeodomain transcription factor CEH-63 that is strongly expressed in the proprioceptive DVC neuron of which the cell body is located in the dorsal-rectal ganglion of the tail. However, laser ablation of DVC or photostimulation of this neuron did not reveal any obvious behavioral changes (Feng et al., 2012).

**Discussion / Future Trends**

Optogenetics has proven useful for optical interrogations of synapses and circuits and the neural basis of behavior. Because of its ability to stimulate or inhibit arbitrary sets of neurons precisely and reversibly, it allows a level of control previously unavailable. In many circuits and systems in *C. elegans*, optogenetic investigations add nuance and complexity to existing models. As the optogenetic toolbox expands and the accuracy of targeted illumination systems continues to improve we expect the usefulness of optogenetics to increase. The combination of optogenetics with optical neurophysiology techniques like calcium imaging (Kerr 2006) and recently available genetically encoded voltage sensors (Kralj et al., 2012; Jin et al., 2012;) will further permit the all-optical dissections of neural networks. Finally, the nascent development of new optogenetic instruments to create virtual sensory environments (Faumont et al., 2011; Kocabas et al., 2012) will create new avenues to study sensory integration and behavior, and brings us closer to achieving fully closed-loop optical neurophysiology investigations whereby optogenetic stimuli would be triggered automatically based on instantaneous readouts of the animal's behavior and its neural activity.

Reference List

Alkema, M.J., Hunter-Ensor, M., Ringstad, N., and Horvitz, H.R. (2005). Tyramine Functions independently of octopamine in the *Caenorhabditis elegans* nervous system. Neuron **46**, 247-260.

Almedom, R.B., Liewald, J.F., Hernando, G., Schultheis, C., Rayes, D., Pan, J., Schedletzky, T., Hutter, H., Bouzat, C., and Gottschalk, A. (2009). An ER-resident membrane protein complex regulates nicotinic acetylcholine receptor subunit composition at the synapse. EMBO J. **28**, 2636-2649.

Ando, R., Hama, H., Yamamoto-Hino, M., Mizuno, H., and Miyawaki, A. (2002). An optical marker based on the UV-induced green-to-red photoconversion of a fluorescent protein. Proc. Natl. Acad. Sci. U.S.A **99**, 12651-12656.

Barclay, J.W., Morgan, A., and Burgoyne, R.D. (2012). Neurotransmitter release mechanisms studied in *Caenorhabditis elegans*. Cell Calcium **52**, 289-295.

Bargmann, C.I., Hartwieg, E., and Horvitz, H.R. (1993). Odorant-selective genes and neurons mediate olfaction in *C. elegans*. Cell **74**, 515-527.

Bargmann, C.I., and Kaplan, J.M. (1998). Signal transduction in the *Caenorhabditis elegans nervous system*. Annu Rev Neurosci **21**, 279-308.

Bargmann C.I., Thomas JH, Horvitz HR (1990) Chemosensory cell function in the behavior and development of *Caenorhabditis elegans.* Cold Spring Harb Symp Quant Biol **55**, 529-538.

Berndt A, Yizhar O, Gunaydin LA, Hegemann P, Deisseroth K (2009) Bi-stable neural state switches. Nat Neurosci **12,** 229-234.

Beverly, M., Anbil, S., and Sengupta, P. (2011). Degeneracy and neuromodulation among thermosensory neurons contribute to robust thermosensory behaviors in *Caenorhabditis elegans*. J. Neurosci. **31**, 11718–11727.

Boulin, T., and Hobert, O. (2012). From genes to function: the *C. elegans* genetic toolbox. WIREs Dev Biol **1**, 114–137.

* Boyden ES, Zhang F, Bamberg E, Nagel G, Deisseroth K (2005) Millisecond-timescale, genetically targeted optical control of neural activity. Nat Neurosci **8**,1263-1268.

Brenner, S. (1974). The genetics of *Caenorhabditis elegans*. Genetics **77**, 71–94.

Busch KE, Laurent P, Soltesz Z, Murphy RJ, Faivre O, Hedwig B, Thomas M, Smith HL, de Bono M (2012) Tonic signaling from O(2) sensors sets neural circuit activity and behavioral state. Nat Neurosci **15**, 581-591.

*C. elegans* Genetics Center, University of Minnesota, http://www.cbs.umn.edu/CGC/

Chalfie M, Sulston JE, White JG, Southgate E, Thomson JN, Brenner S (1985) The neural circuit for touch sensitivity in *Caenorhabditis elegans*. J Neurosci **5**, 956-964.

Chase, D.L., Pepper, J.S., and Koelle, M.R. (2004). Mechanism of extrasynaptic dopamine signaling in *Caenorhabditis elegans*. Nat. Neurosci. **7**, 1096–1103.

Chatzigeorgiou M, Schafer WR (2011) Lateral facilitation between primary mechanosensory neurons controls nose touch perception in *C. elegans*. Neuron **70**, 299-309.

Chatzigeorgiou M, Yoo S, Watson JD, Lee WH, Spencer WC, Kindt KS, Hwang SW, Miller DM 3rd, Treinin M, Driscoll M, Schafer WR (2010) Specific roles for DEG/ENaC and TRP channels in touch and thermosensation in *C. elegans* nociceptors. Nat Neurosci **13**, 861-868.

Cheung BH, Cohen M, Rogers C, Albayram O, de BM (2005) Experience-dependent modulation of *C. elegans* behavior by ambient oxygen. Curr Biol **15**, 905-917.

Chow BY, Han X, Dobry AS, Qian X, Chuong AS, Li M, Henninger MA, Belfort GM, Lin Y, Monahan PE, Boyden ES (2010) High-performance genetically targetable optical neural silencing by light-driven proton pumps. Nature **463**, 98-102.

Croll, N. (1975). Components and patterns in the behavior of the nematode *Caenorhabditis elegans*. Journal of Zoology **176**, 159–176.

Davis MW, Morton JJ, Carroll D, Jorgensen EM (2008) Gene activation using FLP recombinase in *C. elegans*. PLoS Genet **4**, e1000028.

Edwards SL, Charlie NK, Milfort MC, Brown BS, Gravlin CN, Knecht JE, Miller KG (2008) A novel molecular solution for ultraviolet light detection in *Caenorhabditis elegans*. PLoS Biol **6**, e198.

Emtage, L, Aziz-Zaman S, Padovan-Merhar S, Horvitz HR, Fang-Yen C, and Ringstad N (2012) IRK-1 Potassium Channels Mediate Peptidergic Inhibition of *Caenorhabditis elegans* Serotonin Neurons via a Go Signaling Pathway. J Neurosci **32**, 16285–16295.

Erbguth, K., Prigge, M., Schneider, F., Hegemann, P., and Gottschalk, A. (2012). Bimodal Activation of Different Neuron Classes with the Spectrally Red-Shifted Channelrhodopsin Chimera C1V1 in *Caenorhabditis elegans*. PLoS ONE **7**, e46827.

Ezcurra M, Tanizawa Y, Swoboda P, Schafer WR (2011) Food sensitizes *C. elegans* avoidance behaviours through acute dopamine signalling. EMBO J **30**, 1110-1122.

Faumont, S., Rondeau, G., Thiele, T.R., Lawton, K.J., McCormick, K.E., Sottile, M., Griesbeck, O., Heckscher, E.S., Roberts, W.M., Doe, C.Q., et al. (2011). An Image-Free Opto-Mechanical System for Creating Virtual Environments and Imaging Neuronal Activity in Freely Moving *Caenorhabditis elegans*. PLoS ONE **6**, e24666.

Feng H, Reece-Hoyes JS, Walhout AJ, Hope IA (2012) A regulatory cascade of three transcription factors in a single specific neuron, DVC, in *Caenorhabditis elegans*. Gene **494**, 73-84.

Francis, M.M., and Maricq, A.V. (2006). Electrophysiological analysis of neuronal and muscle function in *C. elegans*. Methods Mol Biol **351**, 175-192.

Francis, M.M., Mellem, J.E., and Maricq, A.V. (2003). Bridging the gap between genes and behavior: recent advances in the electrophysiological analysis of neural function in *Caenorhabditis elegans*. Trends Neurosci **26**, 90-99.

Franks, C.J., Murray, C., Ogden, D., O'Connor, V., and Holden-Dye, L. (2009). A comparison of electrically evoked and channel rhodopsin-evoked postsynaptic potentials in the pharyngeal system of *Caenorhabditis elegans*. Invert. Neurosci. **9**, 43–56.

Gao, S., and Zhen, M. (2011). Action potentials drive body wall muscle contractions in *Caenorhabditis elegans*. Proc. Natl. Acad. Sci. U.S.A. **108**, 2557–2562.

Geffeney, S.L., Cueva, J.G., Glauser, D.A., Doll, J.C., Lee, T.H.-C., Montoya, M., Karania, S., Garakani, A.M., Pruitt, B.L., and Goodman, M.B. (2011). DEG/ENaC but not TRP channels are the major mechanoelectrical transduction channels in a *C. elegans* nociceptor. Neuron **71**, 845–857.

Goodman MB, Hall DH, Avery L, Lockery SR (1998) Active currents regulate sensitivity and dynamic range in *C. elegans* neurons. Neuron **20**: 763-772.

Goodman, M.B., Lindsay, T.H., Lockery, S.R., and Richmond, J.E. (2012). Electrophysiological methods for *Caenorhabditis elegans* neurobiology. Methods in cell biology **107**, 409-436.

Gray JM, Karow DS, Lu H, Chang AJ, Chang JS, Ellis RE, Marletta MA, Bargmann CI (2004) Oxygen sensation and social feeding mediated by a *C. elegans* guanylate cyclase homologue. Nature **430**, 317-322.

Guo Z.V., Hart A.C., Ramanathan S (2009) Optical interrogation of neural circuits in *Caenorhabditis elegans*. Nat Methods **6**, 891-896.

Ha, H., Hendricks, M., Shen, Y., Gabel, C.V., Fang-Yen, C., Qin, Y., Colón-Ramos, D., Shen, K., Samuel, A.D.T., and Zhang, Y. (2010). Functional organization of a neural network for aversive olfactory learning in *Caenorhabditis elegans*. Neuron **68**, 1173–1186.

Han, Xue, and Edward S. Boyden (2007). "Multiple-Color Optical Activation, Silencing, and Desynchronization of Neural Activity, with Single-Spike Temporal Resolution." PLoS ONE **2**, e299.

Hedgecock, E.M., and Russell, R.L. (1975). Normal and mutant thermotaxis in the nematode *Caenorhabditis elegans*. Proc Natl Acad Sci USA. **72**, 4061–4065.

Hilliard M.A., Apicella A.J., Kerr R, Suzuki H, Bazzicalupo P, Schafer W.R. (2005) In vivo imaging of *C. elegans* ASH neurons: cellular response and adaptation to chemical repellents. EMBO J **24**, 63-72.

Hilliard M.A., Bargmann C.I., Bazzicalupo P (2002) *C. elegans* responds to chemical repellents by integrating sensory inputs from the head and the tail. Curr Biol **12**, 730-734.

Hilliard MA, Bergamasco C, Arbucci S, Plasterk RH, Bazzicalupo P (2004) Worms taste bitter: ASH neurons, QUI-1, GPA-3 and ODR-3 mediate quinine avoidance in *Caenorhabditis elegans*. EMBO J **23**, 1101-1111.

* Husson SJ, Costa WS, Wabnig S, Stirman JN, Watson JD, Spencer WC, Akerboom J, Looger LL, Treinin M, Miller DM, III, Lu H, Gottschalk A (2012a) Optogenetic analysis of a nociceptor neuron and network reveals ion channels acting downstream of primary sensors. Curr Biol **22**, 743-752.

Husson SJ, Liewald JF, Schultheis C, Stirman JN, Lu H, Gottschalk A (2012b) Microbial light-activatable proton pumps as neuronal inhibitors to functionally dissect neuronal networks in *C. elegans*. PLoS ONE **7**, e40937.

Iseki M, Matsunaga S, Murakami A, Ohno K, Shiga K, Yoshida K, Sugai M, Takahashi T, Hori T, Watanabe M (2002) A blue-light-activated adenylyl cyclase mediates photoavoidance in *Euglena gracilis*. Nature **415**, 1047-1051.

Jin, L., Han, Z., Platisa, J., Wooltorton, J.R.A., Cohen, L.B., and Pieribone, V.A. (2012). Single action potentials and subthreshold electrical events imaged in neurons with a fluorescent protein voltage probe. Neuron **75**, 779–785.

Kang, L., Gao, J., Schafer, W.R., Xie, Z., and Xu, X.Z.S. (2010). C. elegans TRP family protein TRP-4 is a pore-forming subunit of a native mechanotransduction channel. Neuron **67**, 381–391.

Kawano, T., Po, M.D., Gao, S., Leung, G., Ryu, W.S., and Zhen, M. (2011). An Imbalancing Act: Gap Junctions Reduce the Backward Motor Circuit Activity to Bias *C. elegans* for Forward Locomotion. Neuron **72**, 572–586.

Kaplan JM, Horvitz HR (1993) A dual mechanosensory and chemosensory neuron in *Caenorhabditis elegans*. Proc Natl Acad Sci U S A **90**, 2227-2231.

Kerr, R. (2006). Imaging the activity of neurons and muscles. WormBook.

* Kocabas, A., Shen, C.-H., Guo, Z.V., and Ramanathan, S. (2012). Controlling interneuron activity in *Caenorhabditis elegans* to evoke chemotactic behaviour. Nature **490**, 273-277

Koo PK, Bian X, Sherlekar AL, Bunkers MR, Lints R (2011) The robustness of *Caenorhabditis elegans* male mating behavior depends on the distributed properties of ray sensory neurons and their output through core and male-specific targets. J Neurosci **31**, 7497-7510.

Kuhara A, Ohnishi N, Shimowada T, Mori I (2011) Neural coding in a single sensory neuron controlling opposite seeking behaviours in *Caenorhabditis elegans*. Nat Commun **2**, 355.

Kralj, J.M., Douglass, A.D., Hochbaum, D.R., Maclaurin, D., and Cohen, A.E. (2012). Optical recording of action potentials in mammalian neurons using a microbial rhodopsin. Nat. Methods **9**, 90–95.

* Leifer AM, Fang-Yen C, Gershow M, Alkema MJ, Samuel ADT (2011) Optogenetic manipulation of neuroal activitgy in freely moving *Caenorhabditis elegans*. Nat Meth **8**, 147-152.

Lewis JA, Fleming JT, McLafferty S, Murphy H, Wu C (1987) The levamisole receptor, a cholinergic receptor of the nematode *Caenorhabditis elegans*. Mol Pharmacol **31,** 185-193

* Lima SQ, Miesenbock G (2005) Remote control of behavior through genetically targeted photostimulation of neurons. Cell **121**, 141-152

Li W, Kang L, Piggott BJ, Feng Z, Xu XZ (2011) The neural circuits and sensory channels mediating harsh touch sensation in *Caenorhabditis elegans*. Nat Commun **2**, 315.

* Liewald, J.F., Brauner, M., Stephens, G.J., Bouhours, M., Schultheis, C., Zhen, M., and Gottschalk, A. (2008). Optogenetic analysis of synaptic function. Nat Meth **5**, 895–902.

* Lindsay TH, Thiele TR, Lockery SR (2011) Optogenetic analysis of synaptic transmission in the central nervous system of the nematode *Caenorhabditis elegans*. Nat Commun **2**, 306.

Liu, P., Chen, B., and Wang, Z.W. (2011). Gap junctions synchronize action potentials and Ca2+ transients in *Caenorhabditis elegans* body wall muscle. J Biol Chem **286**, 44285-44293.

Liu Q, Hollopeter G, Jorgensen EM (2009) Graded synaptic transmission at the *Caenorhabditis elegans* neuromuscular junction. Proc Natl Acad Sci U S A **106**, 10823-10828.

Macosko EZ, Pokala N, Feinberg EH, Chalasani SH, Butcher RA, Clardy J, Bargmann CI (2009) A hub-and-spoke circuit drives pheromone attraction and social behaviour in *C. elegans*. Nature **458**, 1171-1175.

Mahoney, T.R., Luo, S., Round, E.K., Brauner, M., Gottschalk, A., Thomas, J.H., and Nonet, M.L. (2008). Intestinal signaling to GABAergic neurons regulates a rhythmic behavior in *Caenorhabditis elegans*. Proc. Natl. Acad. Sci. U.S.A. **105**, 16350-16355.

Mellem JE, Brockie PJ, Zheng Y, Madsen DM, Maricq AV (2002) Decoding of polymodal sensory stimuli by postsynaptic glutamate receptors in *C. elegans*. Neuron **36**, 933-944.

Miller KG, Alfonso A, Nguyen M, Crowell JA, Johnson CD, Rand JB (1996) A genetic selection for *Caenorhabditis elegans* synaptic transmission mutants. Proc Natl Acad Sci U S A **93,** 12593-12598

Milward K, Busch KE, Murphy RJ, de Bono M, Olofsson B (2011) Neuronal and molecular substrates for optimal foraging in *Caenorhabditis elegans*. Proc Natl Acad Sci U S A **108**, 20672-20677.

Mori, I., and Ohshima, Y. (1995). Neural regulation of thermotaxis in Caenorhabditis elegans. Nature **376**, 344–348.

* Nagel G., Brauner M., Liewald J.F., Adeishvili N., Bamberg E., Gottschalk A. (2005) Light activation of channelrhodopsin-2 in excitable cells of *Caenorhabditis elegans* triggers rapid behavioral responses. Curr Biol **15**, 2279-2284.

Nagel G., Szellas T., Huhn W., Kateriya S., Adeishvili N., Berthold P., Ollig D., Hegemann P., Bamberg E. (2003) Channelrhodopsin-2, a directly light-gated cation-selective membrane channel. Proc Natl Acad Sci U S A **100**,13940-13945.

* Narayan, A., Laurent, G., and Sternberg, P.W. (2011). Transfer characteristics of a thermosensory synapse in *Caenorhabditis elegans.* Proc. Natl. Acad. Sci. U.S.A. **108**, 9667-9672.

O'Hagan, R., Chalfie, M., and Goodman, M.B. (2005). The MEC-4 DEG/ENaC channel of *Caenorhabditis elegans* touch receptor neurons transduces mechanical signals. Nat. Neurosci **8**, 43–50.

Okazaki A, Sudo Y, Takagi S (2012) Optical silencing of *C. elegans* cells with arch proton pump. PLoS ONE **7**, e35370.

Oren-Suissa M, Hall DH, Treinin M, Shemer G, Podbilewicz B (2010) The fusogen EFF-1 controls sculpting of mechanosensory dendrites. Science **328**, 1285-1288.

Pirri, J.K., McPherson, A.D., Donnelly, J.L., Francis, M.M., and Alkema, M.J. (2009). A tyramine-gated chloride channel coordinates distinct motor programs of a *Caenorhabditis elegans* escape response. Neuron **62**, 526–538.

* Piggott, B.J., Liu, J., Feng, Z., Wescott, S.A., and Xu, X.Z.S. (2011). The Neural Circuits and Synaptic Mechanisms Underlying Motor Initiation in *C. elegans.* Cell **147**, 922–933.

Ramot, D., MacInnis, B.L., and Goodman, M.B. (2008). Bidirectional temperature-sensing by a single thermosensory neuron in C. elegans. Nat. Neurosci **11**, 908–915.

Richmond, J. (2005). Synaptic function. WormBook.

Richmond, J. (2009). Dissecting and recording from the *C. elegans* neuromuscular junction. Journal of visualized experiments: JoVE **24**, 1165

Richmond, J.E. (2006). Electrophysiological recordings from the neuromuscular junction of *C. elegans*. WormBook.

Richmond J.E., Jorgensen EM (1999) One GABA and two acetylcholine receptors function at the C. elegans neuromuscular junction. Nat Neurosci **2**, 791-797

Schmitt C, Schultheis C, Husson SJ, Liewald JF, Gottschalk A (2012) Specific expression of Channelrhodopsin-2 in individual or single neurons of *Caenorhabditis elegans*. PLoS ONE **7**, e43164

Schroder-Lang S, Schwarzel M, Seifert R, Strunker T, Kateriya S, Looser J, Watanabe M, Kaupp UB, Hegemann P, Nagel G (2007) Fast manipulation of cellular cAMP level by light in vivo. Nat Methods **4**, 39-42.

Schultheis C, Liewald JF, Bamberg E, Nagel G, Gottschalk A (2011a) Optogenetic long-term manipulation of behavior and animal development. PLoS ONE **6**, e18766.

Schultheis, C., Brauner, M., Liewald, J.F., and Gottschalk, A. (2011b). Optogenetic analysis of GABAB receptor signaling in *Caenorhabditis elegans* motor neurons. J. Neurophysiol. **106**, 817–827.

Schuske, K., Beg, A.A., and Jorgensen, E.M. (2004). The GABA nervous system in *C. elegans*. Trends Neurosci **27**, 407-414.

Singaram V.K., Somerlot B.H., Falk S.A., Falk M.J., Sedensky M.M., Morgan P.G. (2011) Optical reversal of halothane-induced immobility in *C. elegans*. Curr Biol **21**, 2070-2076.

* Stirman J.N., Crane M.M., Husson S.J., Wabnig S, Schultheis C, Gottschalk A, Lu H (2011) Real-time multimodal optical control of individual neurons and muscles in freely behaving *Caenorhabditis elegans*. Nat Meth **8**, 153-158.

Stirman, J.N., Crane, M.M., Husson, S.J., Gottschalk, A., and Lu, H. (2012). A multispectral optical illumination system with precise spatiotemporal control for the manipulation of optogenetic reagents. Nat Protoc **7**, 207–220.

Sulston, J.E., and White, J.G. (1980). Regulation and cell autonomy during postembryonic development of *Caenorhabditis elegans*. Dev. Biol. **78**, 577–597.

Timbers, T.A., Giles, A.C., Ardiel, E.L., Kerr, R.A., and Rankin, C.H. (2013). Intensity discrimination deficits cause habituation changes in middle-aged *Caenorhabditis elegans*. Neurobiol. Aging **34**, 621–631.

Troemel E.R., Chou J.H., Dwyer N.D., Colbert HA, Bargmann C.I. (1995) Divergent seven transmembrane receptors are candidate chemosensory receptors in *C. elegans*. Cell **83**, 207-218.

Ward, S. (1973). Chemotaxis by the nematode *Caenorhabditis elegans*: identification of attractants and analysis of the response by use of mutants. Proc Natl Acad Sci U S A **70**, 817–821.

Way J.C., Chalfie M. (1989) The *mec-3* gene of *Caenorhabditis elegans* requires its own product for maintained expression and is expressed in three neuronal cell types. Genes Dev **3**, 1823-1833.

Weissenberger, S., Schultheis, C., Liewald, J.F., Erbguth, K., Nagel, G., and Gottschalk, A. (2011). PACα--an optogenetic tool for in vivo manipulation of cellular cAMP levels, neurotransmitter release, and behavior in *Caenorhabditis elegans*. J Neurochem **116**, 616-625.

Wen, Q., Po, M.D., Hulme, E., Chen, S., Liu, X., Kwok, S.W., Gershow, M., Leifer, A.M., Butler, V., Fang-Yen, C., et al. (2012). Proprioceptive Coupling within Motor Neurons Drives *C. elegans* Forward Locomotion. Neuron **76**, 750–761.

White, J.G., Southgate, E., Thomson, J.N., and Brenner, S. (1986). The Structure of the Nervous System of the Nematode *Caenorhabditis elegans*. Philosophical Transactions of the Royal Society of London. Series B, Biological Sciences **314**, 1-340.

WormBase web site, http://www.wormbase.org.

Yoshikawa, S., Suzuki, T., Watanabe, M., and Iseki, M. (2005). Kinetic analysis of the activation of photoactivated adenylyl cyclase (PAC), a blue-light receptor for photomovements of *Euglena*. Photochem. Photobiol. Sci. **4**, 727–731.

Yizhar, O., Fenno, L.E., Prigge, M., Schneider, F., Davidson, T.J., O'Shea, D.J., Sohal, V.S., Goshen, I., Finkelstein, J., Paz, J.T., et al. (2011). Neocortical excitation/inhibition balance in information processing and social dysfunction. Nature **477**, 171–178.

* Zemelman BV, Lee GA, Ng M, Miesenbock G (2002) Selective photostimulation of genetically chARGed neurons. Neuron **33**, 15-22

* Zemelman BV, Nesnas N, Lee GA, Miesenbock G (2003) Photochemical gating of heterologous ion channels: remote control over genetically designated populations of neurons. Proc Natl Acad Sci U S A **100**, 1352-1357

* Zhang F., Wang L.P., Brauner M, Liewald J.F., Kay K., Watzke N., Wood PG, Bamberg E, Nagel G, Gottschalk A, Deisseroth K (2007) Multimodal fast optical interrogation of neural circuitry. Nature **446**, 633-639.

Zhang, Y., Lu, H., and Bargmann, C.I. (2005). Pathogenic bacteria induce aversive olfactory learning in *Caenorhabditis elegans*. Nature **438**, 179–184.

Zheng, Y., Brockie, P.J., Mellem, J.E., Madsen, D.M., and Maricq, A.V. (1999). Neuronal control of locomotion in *C. elegans* is modified by a dominant mutation in the GLR-1 ionotropic glutamate receptor. Neuron **24**, 347–361.